# Ethics of using language editing services in an era of digital communication and heavily multi-authored papers

**George A. Lozano**

Estonian Centre of Evolutionary Ecology

15 Tähe Street, Tartu, Estonia, 50108

dr.george.lozano@gmail.com

# Abstract

Scientists of many countries in which English is not the primary language routinely use a variety of manuscript preparation, correction or editing services, a practice that is openly endorsed by many journals and scientific institutions. These services vary tremendously in their scope; at one end there is simple proof-reading, and at the other extreme there is in-depth and extensive peer-reviewing, proposal preparation, statistical analyses, re-writing and co-writing. In this paper, the various types of service are reviewed, along with authorship guidelines, and the question is raised of whether the high-end services surpass most guidelines' criteria for authorship. Three other factors are considered. First, the ease of collaboration possible in the internet era allows multiple iterations between author(s) and the "editing service", so essentially, papers can be co-written. Second, "editing services" often offer subject-specific experts who comment not only on the language, but interpret and improve scientific content. Third, the trend towards heavily multi-authored papers implies that the threshold necessary to earn authorship is declining. The inevitable conclusion is that at some point the contributions by "editing services" should be deemed sufficient to warrant authorship. Trying to enforce any guidelines would likely be futile, but nevertheless, it might be time to revisit the ethics of using some of the high-end "editing services". In an increasingly international job market, recognizing this problem might prove progressively more important in authorship disputes, the allocation of research grants, and hiring decisions.

English is the language of science. To get their papers ready for submission, scientists of many countries in which English is not the primary language routinely use a variety of manuscript preparation, correction and/or editing services. Journals often suggest that "non-native English speakers" use these services, ignoring the fact that bad scientific writing is a problem that transcends such arbitrary boundaries. For many decades now, the use of these services has been considered standard practice, fully endorsed and often even insisted upon by some national academies and universities in non-English speaking nations. These services vary tremendously in their scope. At one end there is proof-reading and copy-editing, and at the other extreme there is in-depth and extensive peer-reviewing, data analysis, re-writing and co-writing. Proof-reading and copy-editing have long existed, but with the internet facilitating extensive collaboration, high-end services conducted by experts have become increasingly common. Over the past several decades multiauthorship has vastly increased, which in effect has lowered the threshold required for authorship. Given these changes in the way we work, in this paper I ask whether there is a point at which contributions by "editing services" should be deemed sufficient to warrant co-authorship. I conclude that perhaps it is time to revisit the ethics behind some of these long-standing practices.

**Preparation, Correction and Editing Services**

A wide spectrum of preparation, correction and/or editing services is available. The most basic service is proof-reading. A proof-reader is the last line of defence against minor errors that were not caught by the authors, reviewers, editors or copy-editors. A proof-reader corrects unambiguous errors in punctuation, hyphenation, capitalization, word and line spacing, word order, grammar and spelling, and makes queries to the authors about other minor potential problems. In the publishing industry, a proof-reader also compares in painstaking detail an approved version of the manuscript with the galley-proof, which is an exact copy of what is about to be printed in large quantities. Many journals still have proof-readers on staff, but they deal with papers after acceptance, in the final stages before publication. This aspect of a proof-reader's job is not provided by editing services. Their work occurs before the paper is submitted to a journal.

A step up is "copy-editing". A copy-editor corrects some of the same issues that will be covered again later by the proof-reader, but in addition, a copy-editor examines and corrects sentence structure, continuity, verb tense agreement, jargon and journal-specific conventions. Although a copy-editor has more leeway than a proof-reader, the copy-editor's job is to improve the text without making any unnecessary changes. Hence, sentences might be rearranged to allow a better flow of ideas but the writing style will not change too much. For anything more complex, a copy-editor makes queries to the author, for instance, if there are segments that are ambiguous and need clarification. A copy-editor makes the text technically correct, perhaps even structurally perfect, but does not make any significant changes to the content. Most journals have copy-editors on staff, but they deal with papers only after acceptance.

At a higher level, there is "content" or "substantive" editing. This is where the meanings of these terms used by editing services start to diverge from the standard meanings of these terms. Traditionally, the term "content editor" refers to someone does not really address issues that might be covered later by the copy-editor and the proof-reader, but instead focuses on more meaningful issues, hence "substantive". Editors of edited books, for instance, are content editors in the traditional use of the term; they just offer a variety of suggestions. For example, a content editor might suggest that the introduction should be more focused, the methods more detailed, the results rearranged differently and the discussion more closely linked to the introduction and the results. They might get more specific, pointing out segments that are not clear, paragraphs that should be moved or removed, related topics that should be included, etc. However, content editors do not actually write or re-write the manuscript. Among editing services however, substantive editing in this sense is relatively rare.

The "content editors" working for editing services not only make suggestions on what is already there, but also correct, rearrange, expand or delete sentences, paragraphs and whole sections. Functionally, the manuscript is not edited, but rather partially to completely re-written. The extent of the re-writing depends on the service, the expectations, the original state the paper, the willingness of the person working on the paper to delve deeper into it, and, of course, the price. Several iterations between the editing service and the author(s) usually occur, starting with extensive re-writes by both the "editor" and the author(s), and ending with copy-editing and proof-reading. Most editing services, however, do not to use the term "re-write" when describing and advertising their services, but nonetheless, their so-called "substantive editing" services are actually "re-writing" services.

Other services do not worry about such minor points of semantics. For instance, at their "premium" level one service provides "reorganization such that each section includes the proper content (i.e., Results free of methods and interpretation; all general background in the Introduction rather than in the Discussion, etc.)" (Write Science Right 2012). So, essentially, they address the very basics of writing a scientific paper, similar to what a supervisor does when teaching a student how to write his/her first paper(s). Furthermore, at their "superior" level "the editor will spend the additional time needed for major rewriting, reorganization, formatting … This is a good option for clients who want to send a relatively rough draft, who need extensive help with organization, and/or who have multiple figures and tables that are not up to publication quality." (Write Science Right 2012). Quotations are being used in these sections to reduce ambiguity. Note the wording: "clients who want to send a relatively rough draft". This already goes beyond what many supervisors do for their graduate students.

High-level "content editing" is sometimes called "developmental editing". This "service goes much further than a simple language edit – an expert, PhD-qualified, editor with high-impact journal editing experience and proficiency in your field will work on your manuscript in depth." (Macmillan Science Communication 2012). The developmental editor addresses "the question or objective that underpins the research, structural changes required in the manuscript, the coherence and flow of the arguments in the paper, points of ambiguity that require clarification, apparent logical inconsistencies in the stated hypotheses, potential analytical or methodological weaknesses, how well the conclusions appear to be justified by the results, etc." (Macmillan Science Communication 2012). Hence, a developmental editor is an expert in the field, essentially a colleague, whose contribution extends well beyond that of language revisions and effectively re-writes and co-writes the paper with the author(s). Just like in any other multi-authored collaboration, the suggestions that make it to the final version are at the discretion of the primary and/or senior author(s).

Several services offer a rejection and resubmission service, for which, for example, "a project manager will first assess the peer reviewer's comments vis-à-vis the needs of your manuscript, and then customize an editing service package to deliver a manuscript that is consistent with the expectations of the journal. Journal manuscript resubmission involves multiple rounds of editing and close correspondence between the editor and author." (Enago 2012). So, they co-write the manuscript in accordance to the reviewers' demands and produce detailed responses to the reviewers' comments.

Most of the high-level services also offer services called "peer review" and "journal suggestion". "Peer review" does not refer to the peer review that occurs after submission, but rather to a pre-submission peer review, the type that is often conducted "in-house" before submission with the help of colleagues. An expert in the field will review the paper and produce a comprehensive report, pointing out strengths and weaknesses and making a variety of other suggestions, but in this case, the paper will not be re-written. This work is similar to that of a content editor or a pre- or post-submission peer review, in the traditional meanings of the terms. "Journal suggestion" refers to a service whereby experts in the field examine the paper and suggest the most appropriate journals for the paper, depending on the authors' intents and preferences. The only reason to pay for either of these services is if one believes their scientific knowledge of the journals and the field is greater than that of the author(s) and that of nearby colleagues.

Other services start not with the paper, but with the research proposal. Literature reviews can be produced that "carefully take your research objectives and the problem at hand into consideration", "synthesize results into a knowledge base clearly identifying present state of the available knowledge and the unknown information", "identify areas of controversy in the literature" and "formulate questions that need further research." They work "with the inputs provided by the clients and those obtained by our writers after thorough study to prepare a meticulous, persuasive, coherent, clear and compelling research proposal/grant application". They urge potential clients to "provide us with a rough theme of your research; our writers and editors will write a compelling and persuasive research proposal for you" (Manuscriptedit 2012b). There is no ambiguity here: provide "a rough theme" and the writers and editors will do the rest.

The same company, under their "customized services" have three types of services: "customized writing", "data analysis and interpretation" and "specialized consulting". In "customized writing", "the writing is done either from scratch or from a preliminary outline suggested by the client. Our Editors [sic] and writers make a thorough literature search and use the material and methods, and results/raw data provided by the client to write an eloquently presented text." In "data analysis and interpretation" they perform "statistical analysis and interpretation of raw data provided by the clients". This does not refer merely to statistical tests requested by the client, but rather a complete analysis, starting with the raw data and conducting all preliminary analyses, choosing and carrying out the appropriate statistical tests and interpreting the outcome. Finally, the "specialized consulting" service "is for clients who need help with concepts or conceptual content. Our editors can explain scientific concepts and provide feedback about whether your ideas fit with the currently established scientific body of knowledge" (Manuscriptedit 2012a). With these services, it is evident that a deep understanding of the underlying science is not always a pre-requisite to becoming a prolific scientist.

These are just examples of some of the most comprehensive "editing services". As they state themselves, some of these services go far beyond language corrections. They are conducted by specialists within a narrow area of expertise, essentially anonymous colleagues, who contribute extensively to: (1) the conception and design of studies at the proposal stage,

(2) the statistical analysis of the ensuing data, (3) the interpretation in the context of current knowledge and (4) the writing and re-writing of the final product. The purpose here is not to endorse or to criticize any of the services chosen as examples. Dozens and even hundreds of similar services can easily be found online. They are just providing a service and they are not the ones responsible for the papers. The point is that somewhere along this continuum, from mere proof-reading to some of these high-end services, maybe a line is being crossed. It is unclear where the line is, but some of these high end services are clearly conducting work that traditionally has been deemed worthy of authorship.

**What is authorship?**
Authorship of scientific works used to be a relatively simple concept. Since the first scientific journals were created and until early in the 20th century, most papers had a single author. It was easy enough to determine who was responsible, and who should receive credit. That is no longer the case. With the rise in the number of authors per paper, authorship in itself has become a highly debated field of study (Lindsey 1980; Regalado 1995; Drenth 1996; Howard and Walker 1996; Erlen *et al.* 1997; Quencer 1998; Cronin 2001; Hama and Kusano 2001; Rahman and Muirhead-Allwood 2010).

Throughout all authorship guidelines, there is always the caveat that conditions for authorship vary among disciplines. The International Committee of Medical Journal Editors (ICMJE 2012) suggests that "authorship credit should be based on 1) substantial contributions to conception and design, acquisition of data, or analysis and interpretation of data; 2) drafting the article or revising it critically for important intellectual content; and 3) final approval of the version to be published. Authors should meet conditions 1, 2, **and** 3" (my emphasis). The word "and" is a point of contention (Bennett and Taylor 2003) because if it were possible to enforce these guidelines rigidly, many papers would end up having no authors. People working for "editing services" seldom approve the final version (Logdberg 2011), and it has been argued that the third requirement makes it possible to technically adhere to the guidelines while excluding individuals who have otherwise contributed significantly to the study and the manuscript (Matheson 2011). The word "or" would be more realistic, in which case, some of the high-level work conducted by "editing services" would certainly qualify for authorship under criteria 1 and 2.

The Council of Science Editors (CSE) offers similar guidelines, but instead first focus on a set of general principles, indicating, for instance, that determining authorship is not the responsibility of the editors, that all individuals who have "contributed sufficiently" should be listed as authors, and that authors should approve the paper before publication (Council of Science Editors 2012). To actually identify who qualifies as an author, however, they first fall back on the ICMJE guidelines, and then offer their own: "Authors are individuals identified by the research group to have made substantial contributions to the reported work and agree to be accountable for these contributions. In addition to being accountable for the parts of the work …, an author should be able to identify which of their co-authors are responsible for specific other parts of the work. In addition, an author should have confidence in the integrity of the contributions of their co-authors. All authors should review and approve the final manuscript." Other than the point about final approval, these guidelines are substantially different than those of the ICMJE and put a different burden on each of the authors.

The CSE does add that authorship is not appropriate for "professional writers who participated only in drafting of the manuscript and did not have a role in the design or conduct of the study or the interpretation of results". The likely intent here is to prevent people working for "editing services" from being included as authors. However, given the actual nature of the relationship between "editing services" and their clients, in some cases this guideline would have the opposite effect. As described above, the "editing service" sometimes not only writes the papers, but also has a critical role in designing the study and interpreting the results. Finally, the CSE stipulates that authorship is not appropriate for individuals who only provide advice, research space or financial support, which would disqualify many thesis supervisors, lead researchers, group leaders and principal investigators. These guidelines might be well-intentioned, but they do not match the reality of the situation.

The USA National Institutes of Health (NIH 2007) has broader and simpler guidelines: "authorship should be based on a significant contribution to the conceptualization, design, execution, and/or interpretation of the research study, as well as on drafting or substantively reviewing or revising the research article, and a willingness to assume responsibility for the study". In this case, work by high-level editing services is explicitly included, if they were willing to accept responsibility for the study. They are probably never asked.

Many research institutions, national academies, professional societies and journals also have their own authorship guidelines. Sixty percent of biomedical journals (Wager 2007), 53% of science journals, 32% of social sciences journals and 6% of Arts & Humanities journals have authorship guidelines (Bošnjak and Marušić 2012). In contrast, only 11% of professional societies have authorship guidelines in their professional ethics codes (Bošnjak and Marušić 2012). Research and writing are the main requirements for authorship among journals, but two thirds of all ethics codes include research as the sole criterion (Bošnjak and Marušić 2012). Despite this variance, a review and meta-analysis of authorship issues indicates that universally, conceiving the research and/or research design and writing of the manuscript are the most important criteria for authorship (Marušić *et al.* 2011). Not only do guidelines differ, but they only apply to the respective research institutions, national academies, professional societies or journals, so even if they could be enforced, their jurisdictions would be limited.

Other forms of authorship exist that are usually considered to be unethical but are nonetheless fairly common. There are two sides to the coin; instances when individuals who meet the criteria for authorship are not listed as authors, and instances when authorship is given to individuals who do not meet the criteria (Rennie and Flanagin 1994; Bennett and Taylor 2003; House and Seeman 2010; Seeman and House 2010; Wislar *et al.* 2011; E. Smith and Williams-Jones 2012).

"Guest", "honorary" or "gift" authorship refer to instances when someone who would not normally meet the criteria for authorship is nevertheless included as an author. For example, in a survey of USA chemists, about 20% reported having found they were authors of a paper only after it had been printed (Seeman and House 2010). The reasons for inclusion differ. **Guest authorship** refers to the inclusion of senior authors based on the expectation that adding them will enhance the status of the paper during the review process and/or after publication, akin to arriving to a party with a celebrity. Actually, this is a tacit but strong criticism by the authors of the editors, reviewers and eventual readers, who are presumed to be more impressed by the names of the authors than by the quality of the paper (*auctoritas, non veritas, facit legem – authority, not truth, makes the law*). **Gift authorship** is given as part of a gift-exchange arrangement (*quid pro quo*) whereby colleagues and minor contributors who should at most be relegated to the acknowledgments, are elevated to co-authors. **Honorary authorship** is given *ex officio* to senior professors or departmental chairs (*honoris causa*). It is unclear whether the "honour" is genuine or coerced, and if the latter, whether explicitly or tacitly. Thirty-two percent of radiologists report being "asked" to include an "honorary" author in their papers (Eisenberg *et al.* 2011). From 1978 to 1998, the number of authors of research papers in the British Journal of Medicine significantly increased, mostly because of an increase in authorship claims by senior professors and departmental chairs (Drenth 1998).

On the other hand, failure to include as authors individuals who have met the authorship criteria is referred to as "ghost authorship". Ghost authorship occurs when individuals are willingly or unwillingly excluded as authors. Wilful exclusion might occur when it is perceived that including certain authors might decrease the credibility of the paper. For instance, medical studies might be conducted and papers be written by employees of pharmaceutical or instrumentation companies, but authorship might be limited to individuals without any apparent conflicts of interest. The problems and ethics associated with this form of ghost authorship have been discussed at length elsewhere (e.g., Bennett and Taylor 2003; Lacasse and Leo 2010; Matheson 2011; Wislar *et al.* 2011; Dance 2012; Lexchin 2012). A second type of ghost authorship occurs when legitimate authors are unwillingly or unwittingly excluded as authors. In a survey of chemists in the USA, 50% indicated that they had not received a deserved authorship or had not been properly acknowledged (Seeman and House 2010). The problem of potential authors being unwillingly excluded endures because most authorship dispute protocols require disputes to be referred back to the institution where the research was conducted, the place where the senior author usually works, and where the dispute is investigated by the senior author's colleagues. Hence, might makes right. The high-end work of some "editing services" probably falls somewhere in between these two forms of ghost authorship. On one hand, admitting that "editing services" are so deeply involved in a manuscript's production might diminish the credibility of the paper and the reputation of the other authors. On the other hand, offering authorship to the assigned scientist(s) working for the "editing service" is seldom an option, but it has already been argued that people writing for "editing services" should sometimes be included as authors (Jacobs and Wager 2005; Matheson 2011).

To end this authorship quagmire, it has been suggested that journals should switch from naming authors to naming contributors (Rennie *et al.* 1997; R. Smith 1997). This contributorship model is similar to that of movie credits, whereby everyone who contributed to the finished work is mentioned along with their specific role. Most journals ask authors to qualify their contributions to a given paper upon submission, but this information is not included in the paper itself upon publication. Other journals include this information in the paper, but only in addition to the traditional authors' list. So, contributorship has yet to replace authorship. A problem with contributorship is that even assuming contributors are being completely honest, they are not reliable; when asked a second time, only about 30% of authors in a medical journal consistently reported the same contributions in their own papers (Ilakovac *et al.* 2007). Nevertheless, even using a contributorship system, the high-end work conducted by "editing services" would probably have to feature in several prominent roles.

**The proliferation of heavily multi-authored papers**

Despite the increase in the number and prominence of authorship guidelines, they just do not work (Goodman 1994; R. Smith 1997; Eisenberg *et al.* 2011). This is because guidelines are just that, guidelines, not regulations, and enforcement is not possible. One of the principles endorsed by most authorship guidelines is that deciding authorship is not up to the editors, journals or funding agencies, but rather up to the people doing the work. However, co-authors agree only 30% of the time about each others' contributions (Ilakovac *et al.* 2007). Furthermore, as mentioned above, the ICMJE requires that authors meet all three authorship criteria, yet under the condition of anonymity, 60% of radiologists reported that some of the authors of **their own papers** did not meet **any** of the ICMJE authorship criteria (Eisenberg *et al.* 2011). Hence, other than their collective honesty, which has to be weighed against the possibility of infighting and/or collusion among co-authors, there is little preventing all authors from supporting each other's claims of "sufficiently significant" contributions.

The problem is that there is usually no tangible cost to adding more authors, gratuitously or deservedly. Being included as an author usually benefits the one being included, but to the other authors the costs of including one more person are negligible. As long as each author can claim each paper and each citation as his/her own, papers and citations are not being shared by the authors, but rather magically multiplied by the number of authors (Lindsey 1980; Harzing 2010; Põder 2010). A side benefit is that if a paper ever has to be retracted, the more authors there are, the easier it becomes to avoid responsibility (Erlen *et al.* 1997). Hence, as it currently stands, the system rewards increasingly heavily multi-authored papers. As a consequence, over the past 50 years the number of authors per paper has been increasing (Lindsey 1980; Regalado 1995; Drenth 1996; Howard and Walker 1996; Quencer 1998; Cronin 2001). Fifty years ago single authorship was the norm, but these days it is extremely uncommon. The increase in authors is not merely due to greater multi-disciplinary work and research complexity (Papatheodorou *et al.* 2008).

The cost of adding more authors is minimal even when considering intangibles, such as the desire to be referred by name. Papers with one or two authors, which are now rare, are referred using both authors' names, so far so good. Some authors might wish to be first, but it really does not matter. We can think of the classics in any field (Watson and Crick 1953), and usually neither author is considered to be the primary author, but rather the work is usually perceived to be a genuine and equal collaboration. As in the traditional use of the term "author" in literature, credit and responsibility are attributed equally and, more importantly, both authors are mentioned by name. With 3 or more authors, only the first author is mentioned by name, and the rest are just relegated to "*et al.*". Some conflict might exist about the exact order of authorship, but nevertheless, once the 3 author threshold is reached, adding more authors does affect whether other authors are specifically mentioned by name.

When the "*et. al.*" threshold is moved from the third author to the seventh author, once again, the ego effect becomes apparent. The Vancouver referencing style, adopted by the ICMJE, requires all authors to be named in papers with 6 or fewer authors, but for papers with more than 6 authors, only the first 6 are mentioned, and the rest are replaced with "*et al*.". In biomedical journals the frequency distribution of papers with respect to author number increases sharply from one to six authors, but then it drops precipitously at seven and beyond (Epstein 1993). The Vancouver guidelines made it seem like, suddenly, biomedical papers had a new optimum number of authors: 6. The reality is that all authors want their names listed when the paper is cited, so it is okay to have more authors, but in this case only up to a maximum of six.

The trend towards more authors is not slowing down (Rahman and Muirhead-Allwood 2010; Bebeau and Monson 2011; E. Smith and Williams-Jones 2012), and it is not due to greater multi-disciplinary work and research complexity (Papatheodorou *et al.* 2008). Van Loon (1997) argues that there is no justifiable reason why any paper should contain more than 3 authors, and suggests the radical solution that journals should warn authors that papers with more than 3 authors will be less likely to be accepted. This has not happened yet, and it is unlikely to happen any time soon. A less extreme and more practical solution would be to divide each paper and each citation by the number of authors, so credit and impact would be divided among the authors, not multiplied by the number of authors (Engelder 2007; Harzing 2010; Põder 2010). For example, using this method, for the 2001 Nature paper on the initial sequencing and analysis of the human genome, which had 244 authors and received 5968 citations by the end of 2007, instead of each author receiving all 5968 citations, each author would receive his/her fair share of the credit, 5968/244, or about 24 citations (Ioannidis 2008). The trend towards heavily authored papers is self-reinforcing; as the number of authors increases, the relative contribution of each author and the threshold for authorship both decrease. At some point the contribution of each author is so small that relatively, the contribution of "editing services" might be deemed sufficient for authorship.

**The internet**

The authorship inflation was facilitated by technology that makes it easier to communicate, and to collaborate. Before photocopiers became widely available in the 1960s, it was not even feasible to conduct peer reviews (Spier 2002). Instead, editors just chose the best papers, perhaps asked for some minor changes, but mostly accepted papers as submitted. Authors were just not able to collaborate to the degree that is possible today. Even after photocopiers began to be used, actual physical copies had to be made and sent to each co-author, who could then make suggestions, add new paragraphs or re-write existing ones. For the most part, the co-writer was limited to making hand-written notes that the primary author would then incorporate into the next or final version of the manuscript. Times have changed. Now it is even possible for a paper to be co-written simultaneously by several authors located at different parts of the world. Hence, before word processing and email, "editing services" were limited to language correcting, copy-editing and proof-reading; the high-end iterative services of today were just not possible. It is not uncommon for perfectly suitable ethical guidelines to be rendered obsolete by new technologies. In this case, ethical guidelines about the use of "editing services" have just not kept up with the ease of communication and collaboration brought about by the internet.

**Where is the line?**

Ethics differ among cultures. English is the current language of science, and scientists from countries where English is not the primary language are naturally more likely to use "editing services". Originally these might have been strictly English correction services, but over time, and with technology and the internet facilitating the process, these services have expanded to the point it is now possible to use "editing services" to do everything except collect data.

The type of ghostwriting whereby ghostwriters are willingly and knowingly excluded as authors is far more common in medicine than in other fields. In medicine, with lives more directly evidently at stake, there is often more money available, and it might be more important not only be impartial, but to maintain an appearance of impartiality. The availability of funds also makes it possible to use the more expensive "editing services", and to turn ghostwriting into a lucrative and stable career, whereby writers might be satisfied with their pay and not concerned with receiving authorship for their work (Logdberg 2011). These "editing services" can be expensive. Even basic proof-reading and copy-editing can cost about 10 € per "page" (i.e., 250-300 words). The high-end services described above can cost several thousand € per paper. Hence, just a few papers can add a significant burden to most research budgets, a cost that can be prohibitive for all but the best funded research programs. The high cost means that, outside of medicine, use of "editing services" has traditionally been limited to Western and Northern Europe, and Japan. In the last few decades, scientists from other emerging economies, such as China, South Korea and Brazil have begun to take advantage of these "editing services".

Linguistic, financial, and cultural differences notwithstanding, as science becomes more international and inter-disciplinary, these differences are bound to clash. Hiring committees and granting agencies might be interested in scientists with extensive publication and citation records, but they might not consider that candidates might have different attitudes towards "editing services", and might have used them to different degrees. Funding agencies in some jurisdictions might not even allow spending funds on "editing services". Hence, in an international job market, researchers who have relied on "editing services" might see their productivity plummet if they relocate to jurisdictions where "editing services" are not used or allowed. The added cost also means that the cost-effective productivity and impact (Lozano 2010; Zhao and Ye 2011) would be lower even if "editing services" are an allowable expense. In any case, somewhere along the continuum from correcting to co-writing, a line is likely being crossed. The exact location of that line might differ among cultures, but some factors to consider are the number of authors, the degree of interaction between the author(s) and the "editing service", and the editor(s) level of expertise in the subject.

As the number of authors increases, the responsibility, credit and merit of authorship are diluted (Cronin 2001; Bennett and Taylor 2003; Engelder 2007). As the relative contribution of each author decreases, increasingly smaller contributions become worthy of authorship. In some fields and cultures, despite all the guidelines, authorship is extended to technicians, "in-house" editors and reviewers, senior scientists who contributed laboratory space, statisticians, guest and "honorary" co-authors. This process can become self-reinforcing; as more authors are added the threshold for authorship gets even lower, and it becomes easier to justify adding one more author. Consider a paper with $n$ authors ordered according to their relative contributions. As $n$ increases, at a certain point the type of extensive research, analytical and writing work conducted by "editing services" qualitatively and quantitatively might surpass the contribution of the $n^{th}$ author, and hence should be considered sufficient for authorship.

A second factor to consider is the degree of interaction between the "editing service" and the author(s). Before electronic communication was so widespread, repeated cycles of writing, correcting and re-writing between authors and "editing services" were just not feasible. Copy-editors and proof-readers, when used before submission, only corrected the text; they did not add to it or made any substantive changes. Content editors were so in the traditional sense, and offered general comments, without actually re-writing any major sections of the manuscript. However, technology has made it easier to write papers by iterations. It is difficult to define co-authorship, but any genuine collaboration is characterized by repeated interactions between the co-authors as they work on the manuscript (Eggert 2011; Bošnjak and Marušić 2012). As previously described, other than data gathering, "editing services" can be deeply involved with all aspects of producing a paper, from the beginning to the end, starting with a proposal, continuing with developmental editing and ending with copy-editing and proof-reading. It is clear that at least in some cases, the "editing service" and the author(s) essentially co-write the paper.

A third point of consideration is the degree of expertise of the people working for the "editing service". Oddly, a Ph.D., M.D. or someone with similar advanced training would generally not be the best choice to proof-read or copy-edit a paper. This is so for two reasons. First, it would be a waste of their training and expertise. Second, proof-reading in particular, and copy-editing to a lesser degree, require an extreme attention to detail at the exclusion of all else. They require being able to see the trees without being distracted by the forest. This is not to say that proof-reading is not possible for someone with an advanced degree, but such a person would be always tempted to suggest and make more substantive changes. Proof-reading and copy-editing are skills and services that require an eye for detail and knowledge of the language, but they do not require advanced degrees.

Furthermore, even for more comprehensive services such as "content editing", there is no need to use an expert in the particular sub-field. One does not need a deep understanding of what is being said to know whether it is being said in proper English. If the purpose is to correct only the language, one does not need someone who will comment on scientific content. In fact, someone who only has general knowledge of the field would me more likely to recognize unnecessary jargon, to identify sections that are not completely clear, and to make the text more readable for a general audience. Even at the "content editing" level, the work done by "editing services" is similar to that done by thesis supervisors when M.Sc. or Ph.D. students are first learning to write scientific papers, and supervisors are routinely included as co-authors, sometimes for far smaller contributions.

Some of the high end services described above actually guarantee the work will be done by an expert in the particular sub-field, with experience not only publishing in the sub-field, but publishing high impact work in elite journals. Using such a person is overkill to just proof-read or copy-edit a paper, and as discussed above, is not really needed for basic content editing. However, as some of these services clearly explain, such a person is better suited to make other types of changes to the manuscript, such as addressing the literature, expounding the study's rationale, commenting on recently developed techniques, carrying out statistical analyses, and discussing the implications of the study to the field. By most standards, these types of contributions clearly meet and surpass most authorship criteria.

**Conclusions**

It would probably be pointless to endorse only certain companies, or to approve only certain levels of service. As with the authorship or multi-authorship problem, there would be no way of enforcing such guidelines. National funding agencies could prohibit the use of "editing services", which in the long term might improve writing skills, but in the short term would only impede their own country's researchers. Funding agencies could put limits on how much can be budgeted per paper on "editing services", but that would be difficult to regulate. Hence, guidelines about the use of "editing services" would be as effective as general authorship guidelines: completely dependent on people's willingness to follow them. Nevertheless, the first step is to recognize that a potential problem exists.

"Editing services" can substantially contribute to the intellectual content of a paper. Policies allowing the use of "editing services" but disallowing them authorship are incongruous with standard definitions of authorship. Such policies, are, in effect, endorsing ghost authorship, both types of ghost authorship. In an era of increasingly international and multi-authored science, cultural and ethical disputes are bound to arise. However, guidelines just that, guidelines, not regulations, and not only are they unenforceable, but also limited in their respective jurisdictions and scope. Furthermore, guidelines of journals, institutions, professional societies and national academies are often different from and incompatible with each other, so single resolutions to conflicts might not be possible. Should "editing service" authorship disputes ever arise, they ought to be individually investigated taking into context the backgrounds and ethical guidelines relevant to all concerned parties: the journal, the respective institutional and national regulatory bodies, the funding agencies, the "editors", and the scientific cultures of the current and the eventual authors. Finally, hiring and funding committees should be aware of the degree to which international candidates have relied and might have to continue to rely on "editing services". These days, an extensive publication record is no longer predicated on the ability to write.

**Acknowledgements**

I thank colleagues who were kind enough to discuss these issues with me. No proof-reading, copy-editing, substantive editing or co-writing services were used in the production of this manuscript. However, the help of the journal's reviewers, editors and copy-editors is acknowledged and appreciated. I thank the University of Tartu library for giving me access to their online collections. This is publication number 1305 of the ECEE (reg. 80355697).